\documentclass[a4paper]{article}

\usepackage{INTERSPEECH2020}
\usepackage{amssymb,amsmath,epsfig}
\usepackage{textcomp}
\usepackage{amsfonts,amsthm,amsopn}
\usepackage{bm}
\usepackage{url}
\usepackage[capitalise]{cleveref}
\usepackage{graphicx,caption,lipsum}
\usepackage{booktabs,multirow}
\usepackage{placeins}
\usepackage{algpseudocode,algorithm}
\usepackage{color}
\usepackage[table,xcdraw]{xcolor}
\usepackage{bbm}
\usepackage{enumerate}   
\usepackage{adjustbox}
\usepackage[normalem]{ulem}
\usepackage{fixltx2e}

\def\reg{{\rm\ooalign{\hfil
      \raise.07ex\hbox{\scriptsize R}\hfil\crcr\mathhexbox20D}}}

\usepackage{tikz}
\newcommand*\circled[1]{\tikz[baseline=(char.base)]{
    \node[shape=circle, draw, inner sep=0.001pt, 
        minimum height=3pt] (char) {\vphantom{1g}#1};}}

\newcommand{\vct}[1]{\boldsymbol{\mathbf{#1}}} 

\newcommand{\etal}{et al.}

\graphicspath{{./image/}}

\title{Speaker Representation Learning using Global Context Guided \\ Channel and Time-Frequency Transformations}

\name{{Wei Xia, John H.L. Hansen}}

\email{wei.xia@utdallas.edu, john.hansen@utdallas.edu}

\address{
Center for Robust Speech Systems, University of Texas at Dallas, TX 75080 
{\small \tt}}

\begin{document}

\maketitle

\begin{abstract}
In this study, we propose the global context guided channel and time-frequency transformations to model the long-range, non-local time-frequency dependencies and channel variances in speaker representations.
We use the global context information to enhance important channels and recalibrate salient time-frequency locations by computing the similarity between the global context and local features.
The proposed modules, together with a popular ResNet based model, are evaluated on the VoxCeleb1 dataset, which is a large scale speaker verification corpus collected in the wild.
This lightweight block can be easily incorporated into a CNN model with little additional computational costs and effectively improves the speaker verification performance compared to the baseline ResNet-LDE model and the Squeeze\&Excitation block by a large margin.
Detailed ablation studies are also performed to analyze various factors that may impact the performance of the proposed modules.
We find that by employing the proposed L2-tf-GTFC transformation block, the Equal Error Rate decreases from 4.56\% to 3.07\%, a relative 32.68\% reduction, and a relative 27.28\% improvement in terms of the DCF score. 
The results indicate that our proposed global context guided transformation modules can efficiently improve the learned speaker representations by achieving time-frequency and channel-wise feature recalibration.
\end{abstract}


\section{Introduction}
\label{sec:intro}

Automatic Speaker Verification (ASV) task involves determining a person's identity from audio streams. It provides a natural and efficient way for biometric identity authentication. 
Being able to perform text-independent speaker verification that does not utilize any fixed input text content can significantly help us retrieve a target person.  We can use speaker recognition for audio surveillance~\cite{foggia2016audio}, computer access control, and telephone voice authentication for long distance calling~\cite{lee2011joint,crocco2016audio}.
It is also helpful for targeted speech enhancement and separation systems if we have good speaker embeddings~\cite{wang2018voicefilter,vzmolikova2019speakerbeam,kothapally2017speech}. 

Learning a good speaker representation is crucial to speaker verification tasks. The paradigm has shifted from GMM-UBM and factor analysis based methods like i-vector~\cite{matvejka2011full,hansen2015speaker} with a probabilistic linear discriminant (PLDA) back-end~\cite{kenny2010bayesian,prince2007probabilistic} to deep neural network based models. Different neural network architectures~\cite{snyder2018x,michelsanti2017conditional,hajavi2019deep,wana2020multi} were explored to improve the speaker embedding extraction. Margin based softmax loss functions like Angular Softmax~\cite{liu2017sphereface}, Additive Margin Softmax~\cite{wang2018additive}, and recently proposed Additive Angular Margin loss~\cite{deng2019arcface} were useful to learn a more discriminative speaker embedding. Several new temporal pooling methods like attentive pooling~\cite{okabe2018attentive}, Spatial Pyramid Pooling~\cite{jung2019spatial} and GhostVLAD~\cite{xie2019utterance} were presented to aggregate the variable length input features to a fixed-length utterance level representation. Various noise and language robust speaker recognition models~\cite{yu2017adversarial,xia2019cross,mamun2019quantifying,joglekar2019fearless}, and training paradigms~\cite{heigold2016end,Heo:2017ci} have been proposed and significantly improve speaker verification systems' performance.
Cai \etal~\cite{cai2018exploring} introduced a Learnable Dictionary Encoding (LDE) layer to combine frame-level speaker features to an utterance-level speaker embedding.
This ResNet with LDE encoding model has become very successful in various speaker recognition tasks. We use it as the baseline in this study.

Many speaker verification models are based on convolution neural networks, which learn filters to capture local patterns. However, the filter that only operates on the neighboring local context cannot capture long-range, non-local global information. Also, the time-frequency (TF) and channel information at salient regions may not be well emphasized through a standard convolution layer.
Many recent works~\cite{hu2018squeeze,dai2017deformable,bello2019attention,cao2019gcnet,yang2019gated} try to alleviate these issues by improving the encoding of TF and channel information.
One promising approach to accomplish this is a component called the ``Squeeze \& Excitation'' (SE) block~\cite{hu2018squeeze, xia2019sound}, which explicitly models the inter-dependencies between the channels of feature maps. 
Deformable network \cite{dai2017deformable} designs deformable convolution to enhance spatial modeling ability. AAConv~\cite{bello2019attention} uses two-dimensional relative self-attention to augment the convolution operator.


In this study, we introduce a generalized global time frequency context modeling framework for text-independent speaker verification. 
Speech signals contain different information at each time-frequency location. For example, we may pay more attention to high energy parts in the spectrogram.
Our proposed approach tries to better capture long-range time frequency dependencies and channel variances. 
We firstly present the $l_p$ norm based attentive time-frequency context embedding to efficiently model the global speech contextual information. With carefully designed components, the Global Time Frequency Context (GTFC) vector is used for channel and time-frequency wise feature recalibrations.
It aims to get a better combination of the Non-local block~\cite{wang2018non} and SE block~\cite{hu2018squeeze} to adaptively recalibrate the learned feature map and provides time-frequency attention to specific regions. 
Further, we combine the channel wise GTFC and time-frequency GTFC on the score level by a linear fusion. It aggregates the unique properties of each method and makes feature maps more informative on both domains.
We show that with the linear fusion, the Equal Error Rate (EER) of the ASV system decreases from 4.56\% to 2.70\%, a relative 40.79\% reduction. It also has a 38.10\% relative improvement of DCF compared to the baseline ResNet34-LDE model.

In the following sections, we describe the global time frequency modeling framework and corresponding baseline systems in \cref{sec:model}. We provide detailed explanations of our experiments in \cref{sec:exp}, as well as results and discussions in \cref{sec:result}. Finally we conclude our work in \cref{sec:conclusion}.

\begin{figure}[tbp]
    \centering
    \includegraphics[width=0.93\linewidth, height=5.5cm]{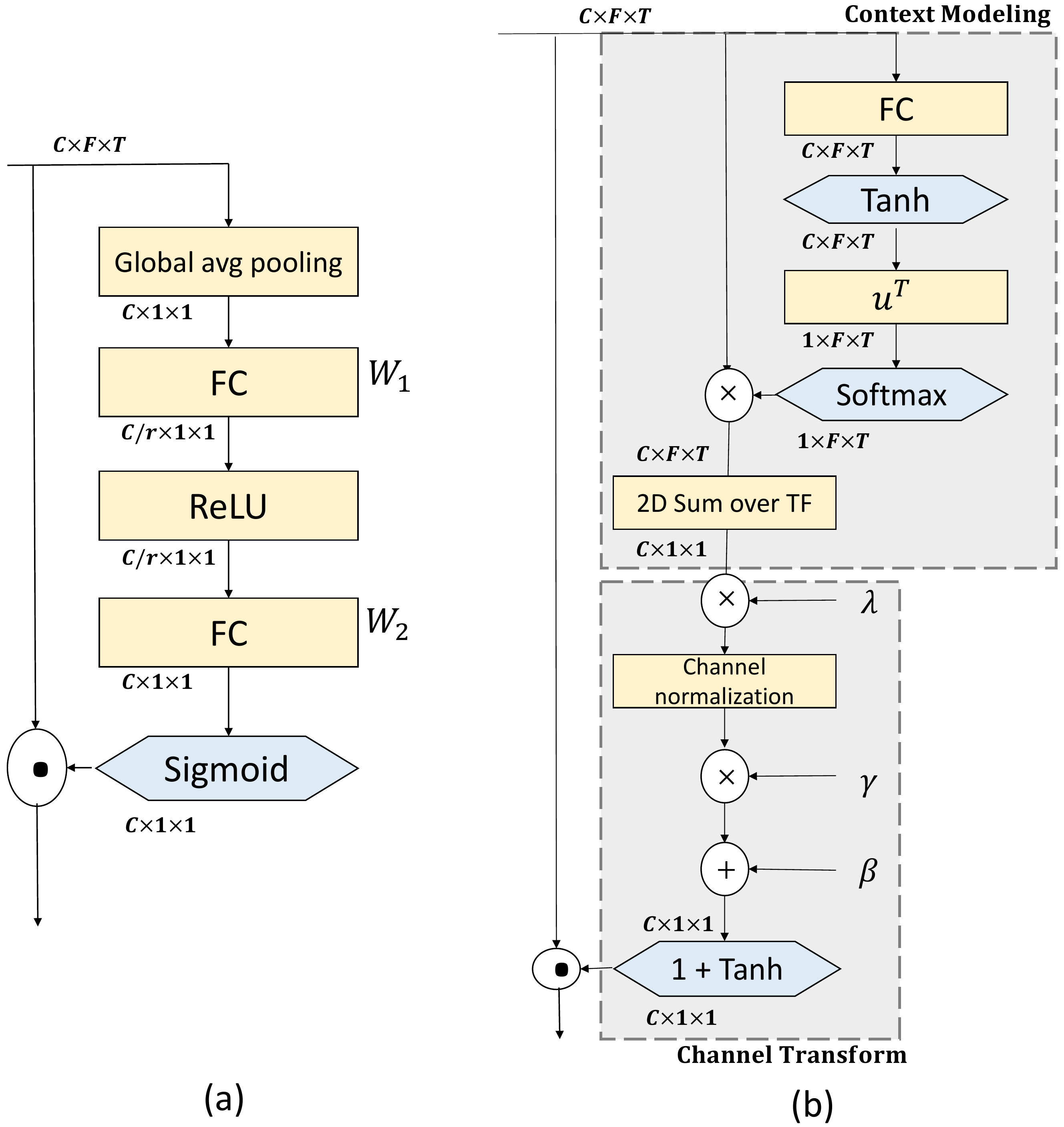}
    \vspace{-1ex}
    \caption{(a) SE block. (b) Proposed global time-frequency context modeling framework and channel-wise transformation.}
    \label{fig:se1}
    \vspace{-2ex}
\end{figure}

\section{Speaker representation learning}
\label{sec:model}

\subsection{ResNet based speaker recognition backbone}
We use the ResNet34 with Learnable Dictionary Encoding (LDE)~\cite{cai2018exploring} as our baseline speaker recognition model.
It uses a well-known ResNet-34 architecture and a Dictionary Learning method to aggregate the variable-length input sequence into a fixed utterance-level speaker embedding.

\vspace{-1ex}
\subsection{Revisiting SE channel attention}

Squeeze and Excitation Network (SE-Net)~\cite{hu2018squeeze} is a well-known method proposed recently to rescale the input feature map to highlight useful channels. Shown in \cref{fig:se1} (a), 
a global average pooling layer is used to generate a channel-wise vector. 
Two fully-connected layers $\mat W_1$ and $\mat W_2$ capture the local channel dependencies. The dimensional reduction factor $r$ indicates the bottleneck in the channel excitation block. 
Finally, with a sigmoid layer, the channel-wise attention vector is obtained to emphasize essential channels. 


\vspace{-1ex}
\subsection{Global time frequency context modeling framework}
The global context information and channel relationships in the SE-Net are inherently implicit. 
To better model long-range relationships and local interactions,
we propose a new generalized framework of global context modeling for channel and time-frequency wise feature recalibration.
We compute an attention map for each time frequency location and attentively pool the corresponding feature values with a $l_p$ norm unit to get the global representation.

We firstly learn a global time frequency embedding $\vct g \in \mathbb{R}^{C}$, then apply channel wise transformation by broadcasting the global TF context vector to each channel; or time-frequency wise transformation by scoring the similarity between the global TF context vector and the local feature vector to get an attention map.


In ~\cref{fig:se1} (b), we show the process to learn the Global Time-Frequency Context (GTFC) embedding and apply it for the channel wise feature map enhancement: (a) the context modeling module groups the features of all positions together via the $l_p$ norm attentive pooling; (b) GTFC is normalized to capture channel-wise dependencies; (c) we use a fusion function to distribute the context vector across channels. In the following, we describe the process in detail, and later in \cref{sec:tf-gtfc} we apply the GTFC embedding for the time-frequency feature enhancement.

\vspace{-1ex}
\subsubsection{$L_p$-norm attentive time frequency context embedding}

A global context embedding module is firstly designed based on $l_p$-norm to aggregate the non-local, long-range time-frequency relationship in each channel. Since individual T-F locations may have different importance, we also use an attention mechanism to focus on salient regions that may have more considerable impact on the global context.
The module can exploit comprehensive contextual information outside small receptive fields of convolutional layers to better encode the global T-F information. Given the embedding weight $\vct \lambda=[\lambda_1,...,\lambda_C]$ along channel and an input feature vector $\vct{x}^{i,j} \in \mathbb{R}^{C}$, the module is defined as the following,
\vspace{-1ex}
\begin{align}
g_{c} = & \lambda_{c} f(\alpha_{i,j},\left\|x_{c}\right\|_{p}) = \lambda_{c}\left\{\left[ \sum_{i=1}^{F} \sum_{j=1}^{T} \alpha_{i,j} |x_{c}^{i, j}|^{p}\right]  \right\} ^{\frac{1}{p}} \\
\alpha_{i,j} & = softmax\left(\vct{u}_{\alpha}^{\mathsf{T}} tanh(\bm{W}_{\alpha} |\vct{x}^{i,j}|^p + \vct{b})\right)
\end{align}
\noindent where $\alpha_{ij}$ is the learned attention weight at a time-frequency location $(i,j)$ through an MLP $\bm{W}_{\alpha}$ and a hidden vector $\vct {u}_{\alpha}$.
The $l_p$-norm unit is efficient at representing nonlinear, complex activations, and is a general form of mean or max pooling with positive values. 
It defines a spherical shape in a non-Euclidean space and summarizes a high-dimensional collection of neural responses. It can avoid the inferior result that average pooling may lead to in some extreme cases.
Additionally, we use an attention mechanism to learn the weight at each time-frequency location for a better global context representation.
We compare the performance of various $l_p$-norms and choose the best one, $l_2$-norm, to be our default setting. 
Trainable parameter $\lambda_c$ is introduced to control the weight of each channel because they may have different significances. 

\vspace{-1ex}
\subsubsection{Channel normalization}
We use a channel normalization method to scale the GTFC vector to capture the competition (high variance) relationship among neuron outputs. 
It is a lightweight operator and reduces the computational cost of the two FC layers used in the SE block from $O(C^2)$ to $O(C)$ but still with a steady performance.
Let $\vct g = [g_1 , ..., g_c ]$, channel normalization is formulated as,
\begin{align}
  \hat{g}_{c}=\frac{k g_{c}}{\|\mathbf{g}\|_{2}}=\frac{k g_{c}}{\sqrt{\sum_{c=1}^{C} g_{c}^{2}+\epsilon}}
\end{align}

\noindent where $\epsilon$ is a small constant for numerical stability. To prevent a too small value of $g_c$ when $C$ is large,  we use a scalar $k$ and set it as $\sqrt{C}$ to normalize the scale of $g_c$, 

\vspace{-1ex}
\subsubsection{Channel gating adaptation}
We finally use a gating mechanism on the normalized GTFC vector to perform channel-wise recalibration on the original feature maps.
Let gating weights $\vct \gamma = [\gamma_1,...,\gamma_C]$ and gating biases $\vct \beta = [\beta_1 , ..., \beta_C]$. They are trainable parameters to adjust the activations of gates channel-wisely.
We use the following gating function with the identity mapping ability.
\vspace{-1ex}
\begin{align}
  \hat{\bm{X}}_{c} = \bm{X}_{c} \cdot \left[1+\tanh \left(\gamma_{c} \hat{g}_{c}+\beta_{c}\right)\right]
\end{align}
The scale of each original feature map $\bm{X}_c \in \mathbb{R}^{F\times T}$ in the channel $c$ is adapted by its corresponding gate so that important channels are emphasized, and less important ones are diminished.
Also, when $\vct \gamma$ and $\vct \beta$ are zeros, we can pass the original features unimpeded to the next layer. This allows any layer to be represented as its initial input. Inspired by ResNet, being able to model the identity mapping can make the network easy to be optimized and robust to the degradation problem in deep networks.
Therefore, we initialize $\vct \gamma$ to and $\vct \beta$ to $0$ in the proposed blocks. 
{\setlength{\belowcaptionskip}{-10pt}
\begin{figure}[tbp]
  \centering
  \includegraphics[width=0.98\linewidth, height=4cm]{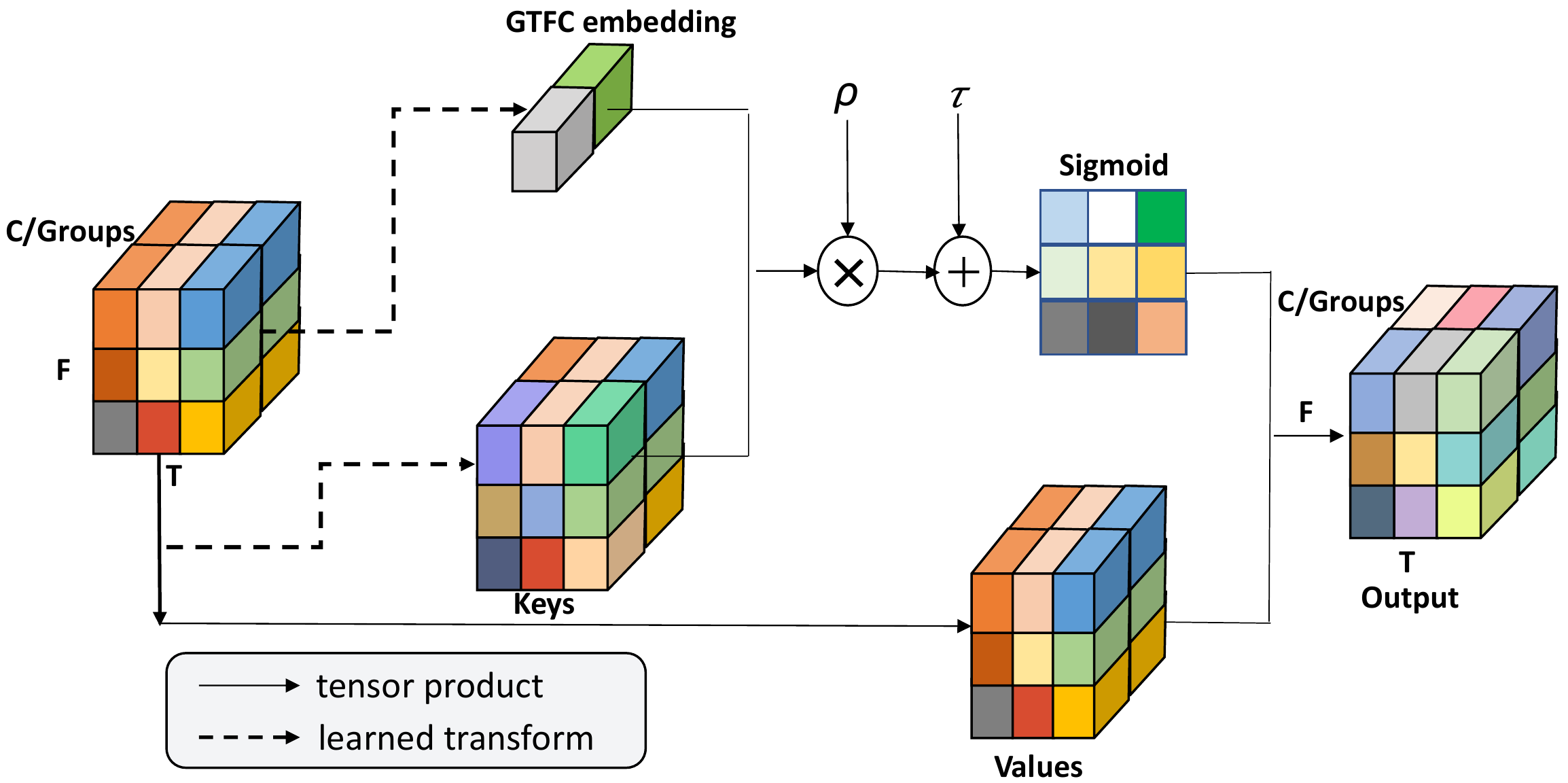}
  \vspace{-1ex}
  \caption{Time-frequency transformation using a group-wise content interaction between the GTFC embedding and the local T-F feature vector.}
  \label{fig:se2}
\end{figure}
}


\subsubsection{Time frequency content interaction}
\label{sec:tf-gtfc}
To capture the time-frequency relationship and analyze which time-frequency location we need to pay attention to, we also propose a way to compute the TF attention map based on the correlation between the GTFC embedding and local feature vectors.
The group wise time-frequency enhancement method is illustrated in~\cref{fig:se2}.
Firstly we divide the $C$ channels, $F\times T$ convolutional feature map into $G$ groups along the channel dimension. We assume that each group could gradually learn a specific response during the training process. In each group, we have a set of local feature vectors $ \mathcal X =\{\vct{x}_{1},...,\vct{x}_{m}\}, \vct{x}_i \in \mathbb{R}^{C/G}, m = F \times T$.
We ideally hope to get features with strong responses at salient time frequency positions (e.g., high energy region). 
However, due to noises and reverberations, we may not be able to get desirable neuron activations after convolution.
To reduce this problem, the group-wise normalized GTFC embedding $\hat{\vct{g}}$ is used as a global group representation, and we compute the correlation between the GTFC vector with the local feature vector $\vct{x}_i$ at each time-frequency location. The similarity score is calculated as the following.
\begin{align}
  e_i & = \text{score}(\hat{\vct{g}}, \vct{x}_i) = \hat{\vct{g}}^{\mathsf{T}} \bm{W}_e \vct{x}_i
\end{align}

With the normalized GTFC vector $\hat{\vct{g}}$, we can generate the corresponding importance coefficient $e_i$ for each position, using the general dot product scoring function~\cite{luong2015effective} in Eq. (5). 
$\bm{W}_e$ is a weight matrice to be learned. In order to prevent the biased magnitude of coefficients between various samples, we also normalize $\vct{e}$ over the time-frequency domain,
\begin{align}
  \hat{e}_i = \frac{e_i - \mu_e}{\sigma_e + \epsilon}, \mu_e = \frac{1}{m}\sum_{j=1}^{m}e_j, \sigma_e^2 = \frac{1}{m}\sum_{j=1}^{m}(e_j - \mu_e)^2
\end{align}
\noindent where $\epsilon$ (e.g., 1e-5) is a constant added for numerical stability. To make sure that the normalization inserted in the network can represent the identity transform, we introduce a pair of parameters ($\rho, \tau$) for each coefficient $e_i$, which scale and shift the normalized value. Finally, to obtain the enhanced feature vector $\hat{\vct{x}}_i$, the original $\vct{x}_i$ is scaled by the generated importance coefficients via a sigmoid function over the space,
\begin{align}
  s_i &= \rho \hat{e}_i + \tau \\
  \hat{\vct{x}}_i &= \vct{x}_i \cdot \sigma(s_i)
\end{align}
All the enhanced features form the recalibrated new feature group. Note that the total number of $\rho$ and $\tau$ is the number of groups, which is negligible compared with millions of model parameters.

\section{Experimental Setup}
\label{sec:exp}

\subsection{Dataset and feature extraction}
To study the effectiveness of the global time-frequency context guided transformation for speaker representation learning, we used VoxCeleb1~\cite{nagrani2017voxceleb} dataset for experiments. It is a large scale text-independent dataset extracted from YouTube videos that contain 153,516 utterances for 1251 celebrities. 
The proposed speaker model is trained only on the VoxCeleb1 development set which contains 1211 speakers. We do not use any data augmentation strategies. There are 40 speakers in the test set with 4874 utterances.
The performance is reported in terms of Equal Error Rate (EER) and minimum Detection Cost Function (DCF) with $P_{target} = 0.01$.

We compute 64 dimensional log-mel filter-bank energies (fbank) on the frame level as input features. A Hamming window of length 25 ms with a 10 ms frame shift is used to extract the fbanks from input audio signals.
We use a random chunk of 300-800 frame features of each audio file as the input to the network, like the strategy used in~\cite{cai2018exploring}.
The input feature is mean and variance normalized on the frame level. Kaldi energy-based VAD is used to remove silent frames. 

\vspace{-1ex}
\subsection{Model training}

    
The baseline model is a ResNet-34 model with LDE pooling~\cite{cai2018exploring} and angular softmax~\cite{liu2017sphereface} (margin $m=4$). We split 90\% of the development set for training, and the remaining 10\% for validation and parameter tuning. 
It is found that the best input feature setting is a frame length of 25 ms, 64 dimensional fbanks. We use the same ResNet34-LDE model parameter settings in ~\cite{cai2018exploring}, 
where residual layers' channel sizes are 16, 32, 64, and 128 respectively.

For our proposed GTFC based models, \textit{Swish}~\cite{ramachandran2017searching} activation function is used at all positions in the ResNet34-LDE model, and we find it is helpful to improve the performance. The model is trained on the VoxCeleb1 training split for 50 epochs with a batch size of 120 on 4 GPUs. 
We use a SGD optimizer with 0.9 momentum and initialize the learning rate as $10^{-3}$ as well as 1e-4 weight decay. The learning rate is reduced by 0.1 when the validation loss does not reduce for 10 epochs. The extracted utterance-level embedding size is 128. $l_2$ norm is used as the default setting in the $l_p$ norm unit.  The general dot-product scoring function is applied to compute the time-frequency attention matrix. We insert the proposed GTFC module after the last Batch Norm layer in each residual basic block. 

For the backend, we use LDA to reduce the dimension of the embeddings to 120, and they are also centered and length normalized. PLDA scoring is applied to evaluate the verification performance.

\section{Results and Discussions}
\label{sec:result}

\subsection{Experimental results}

In order to thoroughly evaluate our proposed methods, we conduct a detailed ablation analysis in this section. We first perform experiments on the GTFC guided channel wise transformation (c-GTFC), followed by the group wise time frequency transformation (tf-GTFC), and then the linear fusion results for speaker verification. 
Finally, we analyze various factors that may affect the performance of GTFC blocks.

From \cref{tab:sed_results}, we observe that our proposed channel wise GTFC block (L2-c-GTFC+ResNet34-LDE)  improves the SV performance by a large margin compared with the ResNet34-LDE model. With the L2-c-GTFC block, overall EER of the ResNet34-LDE model decreases from 4.56\% to 3.13\%, relatively 31.36\%; also from 4.01\% to 3.13\% compared with the SE block, a relative reduction of 21.95\%. It may suggest that our proposed $l_2$ norm based global time-frequency context block can greatly recalibrate the significant feature regions and improve the speaker verification performance.

We also find a consistent performance improvement from L2-c-GTFC model to the L2-tf-GTFC results. The EER reduces from 3.13\% to 3.07\%, with a relative 1.92\% improvement.
It indicates that the L2-tf-GTFC might be more efficient than L2-c-GTFC, which aligns with our assumption that the time-frequency space may have more meaningful information than channels for the SV task.


\begin{table}[htbp]
  \centering
   \caption{SV results on the VoxCeleb1 test set using various models and ResNet34-LDE +our proposed GTFC guided blocks.}
    \begin{tabular}{l|cccc}
        \toprule

       Model   & \multicolumn{1}{c}{EER (\%)} & \multicolumn{1}{c}{DCF}  & \multicolumn{1}{c}{Train Set} \\
    \hline

    Ivector~\cite{nagrani2017voxceleb} & 8.80 & 0.7300 & VoxCeleb1 \\
    Xvector~\cite{okabe2018attentive} & 3.85 & 0.4060 & VoxCeleb1 \\
    UtterIdNet~\cite{hajavi2019deep} & 4.26 & N/A & VoxCeleb2 \\
    SPE~\cite{jung2019spatial} & 4.20 & 0.4220 & VoxCeleb1 \\
    ResNet34-LDE~\cite{cai2018exploring} & 4.56 & 0.4410 & VoxCeleb1 \\
    ResNet34-LDE +SE & 4.01  & 0.3940  & VoxCeleb1  \\
    \hline
    \hline
    +L1-c-GTFC (ours)  & 4.14 &  0.4141 & VoxCeleb1 \\
    +L1-tf-GTFC (ours) & 3.38 & 0.3435 &  VoxCeleb1 \\
    \circled{1} +L2-c-GTFC  (ours) & \textbf{3.13} & \textbf{0.3169}  & VoxCeleb1 \\
    \circled{2} +L2-tf-GTFC (ours) & \textbf{3.07} &  \textbf{0.3207} &  VoxCeleb1 \\
    \circled{1} \& \circled{2} linear fusion & \textbf{2.70} & \textbf{0.2730} & VoxCeleb1\\
    \bottomrule

    \end{tabular}%
  \label{tab:sed_results}%
\end{table}%
\vspace{-1ex}

We further compare the global time frequency context embedding with different $l_p$ norm units. We investigate the $l_1$ norm and $l_2$ norm based GTFC blocks and observe that $l_2$ norm based GTFC blocks perform better than the $l_1$ norm based blocks in all cases. We also tried to set the order $p$ as a learning parameter, but the result is usually worse than the $l_2$ norm, so we use $l_2$ norm based GTFC blocks in all our experiments and subsequent analysis.
Finally, a score level linear fusion with equal weights is employed to combine the L2-c-GTFC and the L2-tf-GTFC results. It achieves the best results with 2.70\% EER and 0.2730 DCF.

\subsection{Empirical analysis}
\textbf{Channel adaptation operator.}
We examine the activation function of the channel gating adaptation in the L2-c-GTFC block with a few different non-linear activation functions and show the results in \cref{tab:ab_results} (a). All the non-linear gating adaptation operators achieve promising performance, and $1 + tanh$ gets the best result. It shows that the identity mapping in the gating function is helpful for the channel adaptation.
\begin{table}[htbp]
  \centering
  \caption{Empirical analysis for different components of our proposed blocks.}
  \begin{adjustbox}{max width=1.0\linewidth}
    \setlength{\tabcolsep}{14pt}
  \begin{tabular}{l|cc}
    \toprule
    \multicolumn{3}{c}{(a) {L2-c-GTFC channel adaptation operator}} \\
    \hline
    Operator & \multicolumn{1}{c}{EER(\%)} & \multicolumn{1}{c}{DCF} \\
    \hline
    sigmoid & 4.69  & 0.4434 \\
    1+ELU & 3.85  & 0.3735 \\
    1+tanh & \textbf{3.13}  & \textbf{0.3169} \\
    \bottomrule
    \toprule
    \multicolumn{3}{c}{(b)  {L2-tf-GTFC normalization parameters}} \\
    \hline
    ($\rho$, $\tau$) & \multicolumn{1}{c}{EER(\%)} & \multicolumn{1}{c}{DCF} \\
    \hline
    (0, 0) & 3.52  & 0.3631 \\
    (0, 1) & \textbf{3.07}  & \textbf{0.3207} \\
    (1, 0) & 4.47  & 0.4729 \\
    (1, 1) & 4.48  & 0.4895 \\
    \bottomrule
    \toprule
    \multicolumn{3}{c}{(c)  {L2-tf-GTFC group number}} \\
    \hline
    Group number & \multicolumn{1}{c}{EER(\%)} & \multicolumn{1}{c}{DCF} \\
    \hline
    4     & 4.69  & 0.4201 \\
    8     & 3.07  & \textbf{0.3207} \\
    16    & \textbf{3.01}  & 0.3451 \\
    \bottomrule 
    \toprule
    \multicolumn{3}{c}{(d)  {L2-tf-GTFC block position}} \\
    \hline
    Block position & \multicolumn{1}{c}{EER(\%)} & \multicolumn{1}{c}{DCF} \\
    \hline
    after BN & \textbf{3.07}  & \textbf{0.3207} \\
    before BN & 3.24  & 0.3718 \\
    before Conv & 3.29  & 0.3885 \\
    \bottomrule
    \end{tabular}
  \end{adjustbox}
    \vspace{-3ex}

  \label{tab:ab_results}%
\end{table}%

\noindent \textbf{Normalization components} $\rho$ and $\tau$.
Shown in \cref{tab:ab_results} (b), we find that the initialization of normalization parameters $\rho$ and $\tau$ in the L2-tf-GTFC block has a considerable effect on the results. Initializing $\rho$ to 0 tends to give better results. With a grid search, we find that the best setting is to assign $\rho$ to 0 and $\tau$ to 1. 
It suggests that in the very early stage of the network training, it might be appropriate to discard the context guided time-frequency attention mechanism. The important thing is to learn a meaningful representation with the convolution stem firstly.


\noindent \textbf{Group number.}
We further investigate the number of groups in the L2-tf-GTFC transformation module in \cref{tab:ab_results} (c).
Too few groups may cause the diversity of feature representations limited.
Using the group number $16$, we obtain the best EER and the group number $8$ for the best DCF values. However, too many groups may also result in a dimension reduction in the feature space, causing a weaker  representation for each group response. We set the group number to $8$ in all our experiments.

\noindent \textbf{Block position.}
Inserting the proposed module after/before the Batch Norm layer, or before the convolution layer in the Residual basic block all improves the results, compared with the baseline ResNet-LDE and SE model. We only insert one proposed block after the Batch Norm layer in our experiments.
The L2-tf-GTFC block only requires about 0.082M  additional parameters,   and therefore is very computationally efficient.

\vspace{0.8ex}
\section{Conclusions}
\label{sec:conclusion}

In this study, we proposed a global time-frequency context modeling framework and successfully applied it to the channel and time-frequency wise feature map recalibration. This model can capture long-range time-frequency dependency and channel variances. With this lightweight block, we can enhance the latent speaker representation and suppress possible distortions. The block was inserted after the last Batch Norm layer of each Residual basic block. The proposed method was evaluated on the VoxCeleb1 dataset, and it was shown to improve the ResNet-LDE and SE models in terms of both EER and DCF by a large margin. Additional analysis and ablation studies indicate that our proposed method can effectively improve the speaker representation learning by strengthening significant time-frequency and channel locations.

\vfill\pagebreak

\newpage

\bibliographystyle{IEEEtran}
\bibliography{main.bib}

\end{document}